\documentclass[twocolumn,aps,prl,showkeywords]{revtex4}

\usepackage[dvips]{graphicx}
\usepackage{amssymb,amsfonts,amsmath,psfrag,latexsym}
\usepackage{graphics}

\def\Q{\mbox{\sffamily\bfseries Q}}

\def\up{\mathbf{u}_{\perp}}
\def\qp{\mathbf{q}_{\perp}}
\def\rp{\mathbf{r}_{\perp}}

\def\bea{\begin{eqnarray}}
\def\eea{\end{eqnarray}}
\def\beq{\begin{equation}}
\def\eeq{\end{equation}}
\begin{document}

\title{Dynamics of a stiff biopolymer in an actively
contractile background: buckling, stiffening and negative dissipation}

\author{Norio Kikuchi$^{1}$, Allen Ehrlicher$^{2}$, Daniel Koch$^{3}$, Josef A. K\"as$^{3}$, Sriram Ramaswamy$^{1,4}$ and Madan Rao$^{5,6}$}
\affiliation{$^1$Centre for Condensed Matter Theory, Department of Physics, 
Indian Institute of Science, Bangalore 560 012, India\\
$^2$Translational Medicine,
Brigham and Women's Hospital, Harvard Medical School,
One Blackfan Circle, Karp 6, Boston, MA 02115, USA\\
$^3$Institute of Soft Matter Physics, Universit\"at Leipzig, Linn\'estra{\ss}e 5, 04103 Leipzig, Germany\\
$^4$Condensed Matter Theory Unit, Jawaharlal Nehru Center for Advanced Scientfic Research, Bangalore
560 064, India\\
$^5$Raman Research Institute, C.V. Raman Avenue, Bangalore 560 080, India\\
$^6$National Centre for Biological Sciences (TIFR),  Bellary Road, 
Bangalore 560 065, India}

\begin{abstract} 
We present a generic theory for the dynamics of a stiff filament under tension, 
in an active
medium with orientational correlations, such as a microtubule in contractile actin. In
sharp contrast to the case of a passive medium, we find the filament can
stiffen, and possibly oscillate, \textit{or} buckle, depending on the contractile or tensile nature of
the activity \textit{and} the filament-medium anchoring interaction. 
We present experiments on the behaviour of
microtubules in the growth cone of a neuron, which provide evidence for these apparently opposing behaviours. We also demonstrate a
strong violation of the fluctuation-dissipation (FD) relation in the effective
dynamics of the filament, including a negative FD ratio. 
Our approach is also of relevance to the dynamics of axons, 
and our model equations bear a remarkable formal similarity to 
those in recent work [PNAS (2001) {\bf 98}:14380-14385] on auditory hair cells. 
Detailed tests of our
predictions can be made using a single filament in actomyosin extracts or
bacterial suspensions.\\
\\
{cytoskeleton $|$ active hydrodynamics $|$ microrheology $|$ contractile $|$ tensile $|$ fluctuation-dissipation ratio  $|$ buckling $|$ stiffening $|$ oscillations $|$ neuronal growth cone $|$ hair cells $|$ axons}\\
\\
Abbreviations: {F-actin, filamentous actin; FD, fluctuation-dissipation}
\end{abstract}

%\pacs{82.35.Lr, 87.10.-e, 87.16.Ka}
\maketitle

\section{Introduction}

The cytoskeleton \cite{alberts} is a dense multicomponent meshwork of semiflexible polymers which interact sterically as well as through active  \cite{alberts,RamaswamyActiveRheo1,curiegrp1} processes. While the blending of polymers industrially requires special effort, the active environment of the living cell provides a setting in which polymers which differ substantially in their stiffness are naturally mixed and interact. Moreover, active processes such as polymerization and the working of molecular motors lead to the generation of stresses without the external imposition of flow fields. These two mechanisms combine to yield a rich range of novel physical phenomena. 
The role of activity in cytoskeletal mechanics is receiving increasing attention, as seen from many recent theoretical and experimental studies of the rheology of cells and cell extracts \cite{LauLubensky03,Mizuno,mack}.
It is clear in particular \cite{multicomprefs} that interactions between different species of filaments are crucial for cell motility, cell division, vesicular transport and organelle positioning and integrity.

\begin{figure}[h]
\begin{center}
\includegraphics[width=7.5cm]{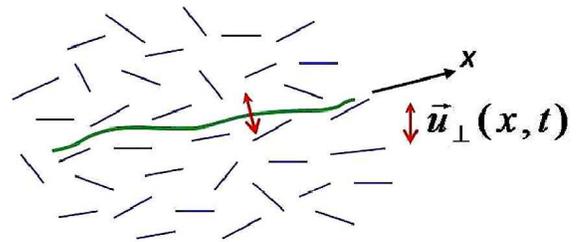}
 \caption{A stiff filament (``microtubule'') embedded in an active isotropic medium consisting of oriented
 fiaments (``F-actin''). The conformations of the microtubule, aligned on an average along the $x$ axis, 
 are described by small  transverse fluctuations $\up\!(x,t)$. The active medium can either be contractile or
tensile. The orientation of F-actin along the microtubule can either be parallel or normal to it.}
      \label{fig1}
      \end{center}
\end{figure}

\begin{figure}[h]
\begin{center}
\includegraphics[width=7.5cm]{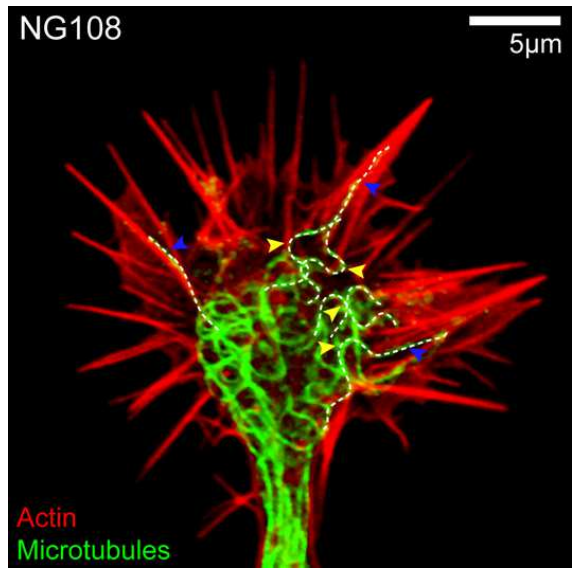}
 \caption{Cytoskeletal polymers in a fixed and fluorescently
stained growth cone of an NG108-15 neuronal cell. In the
peripheral domain actin filaments (red) form dense networks
in the veil-like lamellipodium and bundles in the spike-like
filopodia. Microtubules (green) emerge from the central do-
main into the periphery where they are buckled 
(yellow arrowheads) in the lamellipodium by the actin network 
or stabilized in the filopodial region (blue arrowheads). For illustration some microtubules are traced by
a dotted white line.}
      \label{fig2}
      \end{center}
\end{figure}

In this paper, we make a study of the effect of these interactions by modeling the dynamics of a stiff filament, which we will call a ``microtubule'', immersed in an active medium (Fig.\ref{fig1}) with orientational degrees of freedom, which we will call ``F-actin''.
%\cite{FN1a}
We emphasize here that the names ``microtubule'' and ``F-actin'' are introduced for convenience: we consider both contractile and tensile activity, although only the former applies to actomyosin. Our treatment applies more generally to semiflexible polymers 
under tension in a wide variety of active media.
We describe the medium by the active generalization of liquid-crystal
hydrodynamics
\cite{RamaswamyActiveRheo1,curiegrp1,RamaswamyActiveRheo2,LM031,curiegrp2,LM032,SriramMadanNJP07,curiegrp3,TT95}.
For the purposes of this paper, an active medium is one whose constituent
particles possess the ability to extract energy from an ambient nutrient bath
and dissipate it, executing some kind of systematic motion in the process. This
endows each such particle with a permanent uniaxial stress. The other central
ingredient of this work is \textit{anchoring}, on which we now elaborate. In
general, the interfacial energy of a liquid-crystalline medium at a wall
depends on the relative orientation $\mathbf{n}$ of the molecules of the medium
and the normal $\mathbf{N}$ to the wall. In the simplest cases, it is lowest
for $\mathbf{n}$ parallel, or perpendicular, to $\mathbf{N}$. This interaction
is known as anchoring. In the present work, anchoring enters through the
favoured orientation of the F-actin when confronted with the surface of the
microtubule. We believe that steric or pair-potential effects should lead to
the anchoring of the F-actin normal or parallel to the microtubule. The
interplay of the types of anchoring and activity -- contractile or tensile --
are fundamental to our theory. The filament in our study can be viewed as a
\textit{spatially extended} probe of the active medium, generalizing the
microrheometry of
\cite{LauLubensky03,Mizuno,mack,ChenLauLubensky07,wulibchaber} by simultaneous
access to a wide range of scales. We make contact with earlier work on
oscillatory filaments \cite{hudspeth,pramod2007}, and test our results against
observations on microtubules in a background of contractile actin in neuronal
growth cones (see Fig.\ref{fig2}). 

Here are our main results: 
(i) In the absence of activity, the microtubule
will {\em always} buckle at large anchoring strength, regardless of the type of 
anchoring. (ii) An active medium,
by contrast, can stiffen {\em or} buckle the filament, depending on
the relative signs of activity and anchoring.
A contractile active medium with parallel anchoring always {\em stiffens} 
the filament, as does a tensile active medium with normal anchoring,
while a contractile (tensile) 
medium with normal (parallel) anchoring produces buckling if the strength $W$ of the active stresses is large enough (see Fig.\ref{fig3}). When the nematic correlation length of the F-actin medium is large compared to the linear dimension $L$ of the sample transverse to the microtubule, the buckling wavelength decreases with $L$ as 
$1/\sqrt{WL}$ and $1/\sqrt{W \log L}$ respectively in two and three dimensions. 
(iii) Our observations of the conformations of  
microtubules in the growth cone of neurons are in qualitative accord with these predictions. 
(iv) Activity leads to a breakdown of the fluctuation-dissipation relation: most dramatically, in the regime of strong stiffening, we predict that the effective dissipation turns negative when the frequency crosses a threshold. This is consistent with the observations of \cite{hudspeth} on auditory hair cells and \cite{pramod2007} on axons, and suggests that a negative force-velocity relation at finite frequency should be a generic feature of actively stiffening systems. Indeed, the phenomenological model of \cite{hudspeth} emerges as a limiting case of our fundamental theory. 

\section{A filament in an active medium}

Consider a stiff, locally inextensible filament of total contour length $L$, 
coinciding on average with the $x$ axis (Fig.\ref{fig1}), with unit tangent vector 
%$\mathbf{\hat
%t}\!=\!\mathbf{\hat x}\!+\!\mathbf{\delta{\hat
%t}}\,{\simeq}\,\left(t_x,{\partial}_x\up\!(x,t)\right)$ with
%$t_x\!=\!\sqrt{1\!-\!{\left({\partial}_x\up\right)}{}^2}\,
%{\simeq}\,1+O({\left({\partial}_x\up\right)}{}^2)$, 
$\mathbf{\hat
t}\!=\!\mathbf{\hat x}\!+\!\mathbf{\delta{\hat
t}}\,{\simeq}\,\left(1 + O(\partial_x \up)^2,{\partial}_x\up\!(x,t)\right)$ 
where $\up\!(x,t)$ are small transverse fluctuations, 
and ${\perp}\,{\equiv}\,y,z$, immersed 
in a $d$-dimensional active medium characterized 
%\cite{FN1b} 
by $\Q$, a symmetric traceless nematic order parameter \cite{deGP93}.
The effects of contractile or 
tensile active stresses enter the equations of motion
\bea
{\partial_t}\mathbf{u}_{\perp}\!-\!\mathbf{v}_{\perp}\!(x,\mathbf{r_{\!\perp}}
\!\!=\!\mathbf{0},t)&=& - {1 \over \gamma} \delta F/\delta \up + \!\mathbf{f}_{\perp},
\label{ueqn}\\
{\partial_t}\Q\! 
%-{\lambda + 1 \over 2} \partial_x \mathbf{v}_{\perp}-
%{\lambda - 1 \over 2} \nabla_{\perp}{v}_{x}
\! &=& -{1 \over \zeta}\delta F/\delta \Q 
+\mbox{\boldmath ${\eta}$},\label{Qeqn} 
\eea
%\end{widetext}
for $\up$ and $\Q$ through the hydrodynamic velocity field $\mathbf{v}$ whose dynamics is governed by Eq.\ref{veqn} below. 
The Gaussian, spatiotemporally white noises $\mathbf{f}_{\perp}$, \mbox{\boldmath ${\eta}$} in Eqs.\ref{ueqn},\ref{Qeqn} have strengths $2N_1$, $2N_2$, reducing 
respectively to $2k_BT /\gamma$, $2k_BT/\zeta$ for the equilibrium case. 
In Eqs.\ref{ueqn},\ref{Qeqn}, $\up$ and $\Q$ are 
coupled only through 
the free-energy functional 
$F[\up,\Q]= F_f[\up]+F_{LD}[\Q]+F_{anc}[\up,\Q]$.
%$F[\mathbf{\hat t},\Q]= F_f[\mathbf{\hat
%t}]+F_{LD}[\Q]+F_{anc}[\mathbf{\hat t},\Q]$. 
The filament
free energy $F_f[\up]=\int_0^L\!\!dx[({\sigma}/{2})\,(\partial_x \up)^2
+({\kappa}/{2})(\partial_x^2 \up)^{2}]$
contains bending energy with rigidity $\kappa$ and an imposed
tension 
$\sigma$.
\cite{HFK05,HFK07} to leading order in $\partial_x \up$. The Landau-de
Gennes free energy
$F_{LD}[\Q]=\int\!\!dx\!\!\int\!\!d^2r_{\!\perp}[({a}/{2}){\Q}{}^2+({K}/{2})
{(\nabla\Q)}{}^2]$
describes incipient orientational ordering in the medium \cite{deGP93}.  
We work here in the isotropic phase with correlation length ${\sim}\sqrt{K/a}$; 
a study of the nematic phase by generalising \cite{Yodhfootnote} to include activity will be presented elsewhere \cite{nkikuchi08}. 
%but near the isotropic to nematic
%transition point for $K{\gg}\,a(>\!\!0)$ or far from it for $a{\gg}K$. Note our model
%do not capture the nematic phase where higher order $\underline{Q}$ terms are
%necessary. 
The filament anchors the orientational degrees of freedom of the medium through 
\bea
&F_{anc}[\up,\Q]  = \frac{A}{2} \int_0^L dx \hat{\bf t} \cdot \Q (x,\mathbf{r}_{\perp}=\mathbf{0})\cdot\hat{\bf t} &\nonumber \\
&  \,\,\,\,\,\,\,\,\,\,  \simeq \mbox{const.} + A\!\!\int_0^L\!\!dx 
[\partial_x \up \cdot \Q_{x \perp}(x,\mathbf{r}_{\perp}=\mathbf{0}) +O{\left({\partial}_x\up\right)}{}^2]&
\nonumber
\eea
with negative and positive $A$ corresponding respectively to parallel 
and normal anchoring. 
Note that Eq.\ref{ueqn} generalises \cite{HFK05,HFK07} to 
include anchoring and hydrodynamic flow. In Eq.\ref{Qeqn}, we ignore 
flow-orientation coupling terms \cite{deGP93}
\footnote{If included these would lead to shifts of effective Frank constants and additional possible instabilities in the effective 
equation of motion Eq.\ref{ueqnFT}.}.

For thin-film samples at large F-actin concentration, as in the case of the lamellipodium of adhering cells, it is appropriate to treat the hydrodynamic velocity field in a local-friction approximation. We therefore write $\Gamma{v_i}(x,\mathbf{r_{\!\perp}},t)= -\nabla_j\sigma_{ij}$,
%at low Reynolds number
with $\Gamma \sim \mu/\ell^2$ where $\mu$ is the cytoplasmic viscosity, and the screening length \cite{muthuedwards} $\ell$ is no larger than the film thickness. We ignore pressure gradients on the assumption that the film thickness adjusts to accommodate these. The crucial piece of the stress $\sigma_{ij}$ is the active contribution $\sigma_{ij}^{act}\simeq Wc_0Q_{ij}(x,\mathbf{r_{\!\perp}},t)$
%  \cite{FN2a} 
\footnote{Passive stresses arising from the free-energy functional enter only at higher order in gradients.}
\cite{RamaswamyActiveRheo1,curiegrp1,RamaswamyActiveRheo2,LM031,LM032,SriramMadanNJP07}, where $W <0$ and $W>0$ respectively correspond to contractile and tensile stresses, and 
$c_0$ is the mean F-actin concentration.
To leading order in gradients and linear order in filament undulations, the velocity transverse to the microtubule is 
%\bea
\beq
\label{veqn}
%{v_x}(x,\mathbf{r_{\!\perp}},t)&=&-(Wc_0/\Gamma)\nabla_{\perp}\cdot\Q_{x\perp}
%\!(x,\mathbf{r_{\!\perp}},t)\nn
{\mathbf{v}_\mathbf{\perp}}(x,\mathbf{r_{\!\perp}},t)=-(Wc_0/\Gamma)\partial_x\Q_{x\perp}\!(x,\mathbf{r_{\!\perp}},t).
%\eea
\eeq
%The flow-alignment terms with
%coefficient $\lambda\pm{1}$ are standard in nematic hydrodynamics ($\lambda$:
%reversible kinetic coefficient) \cite{deGP93} (see also
%\cite{RamaswamyActiveRheo,Rheochaos}). The symmetric and anti-symmetric parts describe
%the response of the nematic director to the shear and the rotation of the director %with fluid without other forces. 
%spatiotemporally delta-correlated,
%$\left<f_{\perp{i}}(x,t)f_{\perp{j}}(x^{\prime},t^{\prime})\right>=2N_1\delta_{ij}\delt%a(x-x^{\prime})\delta(t-t^{\prime})$
%with effective temperature $N_1\gamma$. Note, however, that a noise
%results from the non-equilibrium nature of the active medium with property
%$\left<\eta_i(\mathbf{r},t)\eta_j(\mathbf{r}^{\prime},t^{\prime})\right>=2N_2\delta_{ij%}\delta^d(\mathbf{r}-\mathbf{r}^{\prime})\delta(t-t^{\prime})$.
%Here, $N_2\zeta$ is defined to be the "activity temperature". In thermal
%equilibrium, $N_1\gamma=N_2\zeta=k_BT$. For simplicity we used delta-correlated
%noise.
%%%%%%%%%%%%%%%%%%%%%%%%%%%%%%%%%%%%%%%%%%%%
%{\it Results}: 
From Eqs.\ref{ueqn}-\ref{veqn}, the effective Fourier-transformed equation of motion for $\up(q_x,\omega)$ is
%We now analyze the mode structure of the effective dynamics of the filament in the %active medium to study the stability/instability at long wavelength. 
%$\mathbf{q}\!\rightarrow\!{0}$. 
%This is achieved by Fourier transforming (\ref{ueqn}),
%\begin{widetext}
\beq
\label{ueqnFT}
\left(-i\omega+{\sigma \over \gamma}q_x^2+{\kappa \over \gamma}q_x^4\right)
\up
=iq_x\alpha\Q_{x\perp}\!(\mathbf{r_{\!\perp}}\!\!=
\!\mathbf{0})+\mathbf{f}_{\perp},
\eeq
where $\up$ is coupled to $\Q$ only at
$\rp = 0$.
\beq
\label{QeqnFT}
\Q_{x\perp}\!(\mathbf{r_{\!\perp}}\!\!=\!\mathbf{0})=
{iq_x\beta{\zeta} \over K}\!\!\!\int_{\mathbf{q_{\perp}}}
\!\!\!\!\!\!G_{\mathbf{q}\omega}\up
+{\zeta \over K}\!\!\!\int_{\mathbf{q_{\perp}}}
\!\!\!\!\!\!G_{\mathbf{q}\omega}
\mbox{\boldmath ${\eta}$}(q_x,\mathbf{q_{\perp}},\omega),
\eeq
%\end{widetext}
where $q_{\omega}^2{\equiv} - {\zeta}i\omega/K+a/K+q_x^2$ and $G_{\mathbf{q}\omega} \equiv (q_{\!\perp}^2+q_{\omega}^2)^{-1}$, and 
$\int_{\mathbf{q_{\perp}}}\!\!\!{\equiv}\!\!\int_0^{\Lambda}\!\!{d^{d\!-\!1\!}(q_{\!\perp}/2 \pi)}$, with ultraviolet cutoff $\Lambda \sim$ (filament thickness)$^{-1}$.
The signs of $\alpha \equiv (A/\gamma-Wc_0/\Gamma)$ and $\beta \equiv -A/\zeta$ decide the fate of the filament (stiffening \textit{or} buckling) in the following analysis. 
%
%and ${\hat
%q_{\!\perp}^2}=q_{\!\perp}^2$ in $2d$, ${\hat
%q_{\!\perp}^2}=q_y^2+(K/K^{eff})q_z^2$ for $Q_{xy}$ and ${\hat
%q_{\!\perp}^2}=q_z^2+(K/K^{eff})q_y^2$ for $Q_{xz}$ in $3d$, and
%
%$\mathbf{q_{\perp}}$ integral in (\ref{QeqnFT}) is easily performed, 
Since $\int_{\mathbf{q_{\perp}}} G_{\mathbf{q}\omega} 
\sim \ln(\Lambda^2/q_{\omega}^2)$ in $3$d and 
$
%\pi/(2q_{\omega})\!-\!1/\Lambda\simeq
\pi/(2q_{\omega})$ in $2$d, 
%. Note that instead of the logarithmic term in $3$d, we have the $1/q_{\omega}$ 
%dependence of $\mathbf{Q}_{x\perp}$ in $2$d: the effect of $q_{\omega}$ is stronger. 
%Substituting resulting $\mathbf{Q}_{x\perp}
%\!(q_x,\mathbf{r_{\!\perp}}\!\!=\!\mathbf{0},\omega)$ into (\ref{ueqnFT}) 
we obtain the dispersion relations for $\up\!(q_x,\omega)$ for the case of an unbounded medium:
\bea
-i\omega=\left\{
\begin{array}{cl}
-\left[{\sigma \over \gamma}+{\alpha\beta\zeta \over K}
\ln\left({\Lambda^2}/{q_{\omega}^2}\right)\right]q_x^2 + O(q_x^4),& d=3 \\
-\left[{\sigma \over \gamma}+
{\alpha\beta\zeta \over 4K}q_{\omega}^{-\!1}\right]q_x^2 +O(q_x^4),& d=2  .\label{disprel}
\end{array}
\right.
\eea
The case of a confined medium is discussed towards the end of the paper.

\section{Active stiffening and buckling}

\begin{figure}[h]
\begin{center}
\includegraphics[width=9.0cm]{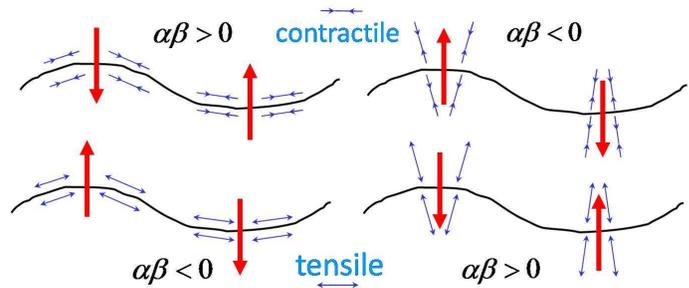}
 \caption{A perturbation of a ``microtubule'' (black) leads, through anchoring, 
to a distortion of the ``F-actin'' medium and hence of the active stress profile. Depending on whether the actin is anchored parallel or normal to the microtubule, and whether the active stresses (blue double-arrows) are contractile or tensile along the actin filaments, the resulting secondary flows (red arrows) either suppress {\em or} enhance the perturbation, leading to active stiffening or buckling.}
      \label{fig3}
      \end{center}
\end{figure}

We are now in a position to investigate the (in)stability of the filament.
%Since
%all parameters inside the braket [$\,\,$] are positive except for $\alpha$ and
%$\beta$, only these parameters can alter the sign of inside the braket hence
%determine the (in)stability. 
At thermal equilibrium ($W=0$) sgn$[\alpha] = - $sgn$[\beta]$; indeed, $\alpha \beta = -A^2/\zeta \gamma <0$ irrespective of the sign of $A$. This implies a buckling instability  of the filament as the anchoring strength is increased, regardless of whether the anchoring is parallel or perpendicular. This is consistent with observations \cite{Yodhfootnote} of buckling of filaments in isotropic solutions of \textit{fd} virus (see their Fig. 1a).
When activity $W$ is switched on, the signs of $\alpha$ and $\beta$ become independent. For large enough $|W|$, $\alpha\!>\!0$ for contractile ($W<0$) activity and $\alpha < 0$ for tensile ($W >0$) activity, while $\beta$ is separately controlled by the nature of the anchoring (Fig.\ref{fig3}). 

Defining the correlation length $\xi \equiv \sqrt{K/a}$ of the F-actin medium, we focus on two limiting cases: (I) deep in the isotropic phase; (II) close to the transition to a nematic phase.
%, and $\alpha\!<\!0$ for tensile activity, $W\!>\!0$. In thermal equilibrium, there is %no activity $W=0$, hence $\alpha=A/\gamma$. The sign of $\beta=-A/\zeta$
%is $\beta\!>\!0$ for parallel and $\beta\!<\!0$ for normal anchoring.
In (I), $q_x \xi \ll 1$, 
%$a/K \gg \left|-{\zeta}i\omega/K+q_x^2\right|$, 
so Eq.\ref{disprel} holds with $q_{\omega} = \sqrt{a/K}$ for small $\omega$. 
%\newline(I) $a/K \gg \left|-{\zeta}i\omega/K+q_x^2\right|$: 
%Expanding $q_{\omega}$ and solving Eq.(\ref{disprel}) for $\omega$ yields, to 
%leading order,
%\bea
%-i\omega=\left\{
%\begin{array}{cl}
%-\left[{\sigma}/{\gamma}+({\alpha\beta\zeta}/{K})
%\ln\left({\Lambda^2}K/a\right)\right]q_x^2,& \quad (3\mbox{d})\\
%-[\sigma/\gamma+\alpha\beta\zeta/(4K\sqrt{a/K})]q_x^2, & \quad 
%(2\mbox{d}).\label{disprelIso}
%\end{array}
%\right.
%\eea
%Note that we have another mode $-i\omega=-2a/\zeta$ in $2$d, yet it decays away much
%faster than the dominant mode in (\ref{disprelIso}).
%We derive interesting (striking?) results.
Thus, for a filament in a strongly active medium, contractile (tensile) activity with parallel (normal) anchoring ($\alpha\beta>0$) leads to enhanced tension, i.e., stiffening.
On the other hand, for strong contractile (tensile) activity with normal (parallel) anchoring ($\alpha\beta<0$) the filament becomes unstable to buckling.
This is explained graphically in 
(Fig.\ref{fig3}). 
%%%%%%%%%%%%%%%%%%%%%%% Figures %%%%%%%%%%%%%%%%%%%%%%%%%%%%%%%
%%%%%%%%%%%%%%%%%%%%%%%%%%%%%%%%%%%%%%%%%%%%%%%%%%%%%%%%%%%%%%%
In (II), $q_x \xi \gg 1$ (which should be accessible in F-actin, see \cite{Viamontes})
so that a wide range of modes of $\Q$ contribute in Eq.\ref{QeqnFT}. 
%$a/K \ll \left|-{\zeta}i\omega/K+q_x^2\right|$,
In dimension $d=$3, Eq.\ref{disprel} applies with $q_{\omega}^2 \simeq q_x^2 -\zeta{i}\omega/K$, so that the dispersion relation differs from case (I) only by a logarithmic factor. In $d=2$, however, solving the resulting cubic equation we find 
\beq
-i\omega=
Y(|\alpha\beta|/4)^{2/3}(\zeta/K)^{1/3}q_x^{4/3}
\label{disprelIN}
\eeq
where $Y=(-1/2{\pm}i\sqrt{3}/2)$ for $\alpha\beta\!>\!0$ and $1$ for $\alpha\beta\!<\!0$. 
Thus, in $2$d, close to the ordering transition of the F-actin, while stability is determined by the sign of $\alpha\beta$ as before, the stable stiffening case $\alpha\beta\!>\!0$, which can arise only for an active system, shows a damped \textit{oscillatory} response, which should appear as dispersive propagating waves on the filament, with speed $\propto q_x^{1/3}$. In the unstable case, the buckling wavelength can be found by comparing the 
negative tension and bending elasticity terms. For an F-actin medium of finite lateral extent $L$, scaling arguments applied to Eqs.\ref{disprel},\ref{disprelIN} lead to a buckling wavelength varying as $1/\sqrt{WL}$ ($d=2$) and $1/\sqrt{W \log L}$ ($d=3$) as  
claimed earlier in the paper.

The ideas presented above are best tested in a model system in which both parallel and normal alignment of F-actin on microtubules are naturally realised.  
The growth cones of neurons provide a convenient system for this purpose, and allow a demonstration of activity-induced stiffening as well as buckling. Our observations on this system are qualitative but in clear accord with the predictions of this paper. 
As seen in Fig.\ref{fig2} microtubules are strongly buckled in the wide central region known as the lamellipodium and show a stabilized straight shape in the spiky protrusions called the filopodia. As we have remarked, purely equilibrium effects in most systems generally lead to buckling. 
It seems, however, that the dynamic activity within the lamella may contribute to microtubule stiffening
in the periphery, as well as buckling in  the central region.
We suggest the following mechanism : when the growing microtubule enters the lamellipodium, it meets the actomyosin network, with actin filaments pointing in a variety of directions. The typical encounter will have a substantial angle between microtubule and actin filament. The analysis above would then predict that the microtubule should buckle, as it does. Upon entering a filopodial region, the microtubule meets actin aligned parallel to it and is stiffened, which again is consistent with our mechanism. 
However, whether the relevant activity is entirely non-specific and based on contractility, or whether
specialized proteins are involved remains an open question.
In another relevant study, Brangwynne \textit{et al}. \cite{mack} observed significant bending fluctuations of microtubules in reconstituted actomyosin networks (see Fig.1 in \cite{mack}). Since the actomyosin is contractile, our theory would predict that the contacts between actin and microtubule are at near-normal alignment, and we look forward to independent tests of this. Indeed, our work provides a theoretical justification for the pointlike normal force that \cite{mack} suggest is responsible for the buckling. 

\section{Fluctuation-Dissipation ratio and negative dissipation}
If $\chi(t)$ is the displacement of a degree of freedom of a system at time $t$ in response to an impulsive force at time $0$, and $C(t)$ is the time-correlation of spontaneous fluctuations in that degree of freedom, the Fluctuation-Dissipation Theorem (FDT) \cite{kubobook2}, which applies to systems \textit{at thermal equilibrium}, says $C(t) = -k_B T \chi(t)$, where $k_B$ is Boltzmann's constant and $T$ is the temperature. This profound and universal connection can be understood in the familiar context of a particle undergoing Brownian motion in a fluid, where both the damping of an initially imposed velocity and the random motion of the particle when no velocity is imposed arise from the same microscopic collisions with molecules. 
For equilibrium systems, this allows one to obtain transport quantities such as conductivity without drawing a current. In systems far from equilibrium, but where the bath producing the fluctuations is still thermal, a judicious definition of variables \cite{speck} can resurrect the FDT. For more general nonequilibrium systems too, a comparison of correlation and response can sometimes offer a useful notion \cite{kurchan} of an effective temperature. Our model system shows radical departures from such benign behaviour.

We restrict attention to parameter ranges where there is no instability, so a steady state exists. Let $S({q_x},\omega)\!\equiv\!
\int_{x,t}\exp(i q_x x - i \omega t) \left<\up\!(x,t)\!\cdot\!\up\!(0,0)\right>$ 
be the correlation function and $\chi^{\prime \prime}(q_x,\omega)$ the imaginary part of the response to an external force $\mathbf{h}(x,t)$ coupled to $\up$ via a term $-\int dx \mathbf{h}\cdot\up$ in the free-energy functional $F$ in Eq.\ref{ueqn}. The departure from unity of the fluctuation-dissipation (FD) ratio $R(q_x,\omega) \equiv (\omega /2 k_BT)S(q_x,\omega)/\chi^{\prime\prime}(q_x,\omega)$ 
is a quantitative measure of nonequilibrium behaviour. We find not only that $R(q_x,\omega)$ depends on its arguments but that it can turn negative for the stable stiffening case $\alpha\beta\!>\!0$. The calculation, from 
Eqs.\ref{ueqnFT},\ref{QeqnFT}, is straightforward. We find 
\beq 
\label{respimag}
\chi{\prime\prime}(q_x,\omega)=\omega \left[{1-\alpha\beta\left(\zeta/\!K\right)
\!{}^2q_x^2\Sigma(q_{\omega}) \over {\cal D}_{q_x\omega}}\right], 
\eeq
where ${\cal D}_{q_x\omega}$ and  $\Sigma(q_{\omega})=\!\!\int_{\mathbf{q_{\perp}}}\!\!\!\left[\!{\left(q_{\!\perp}^2\!+\!a/K\!+\!q_x^2\right)}{}^{\!2}\!\!+\!(\zeta\omega/K)^2\right]^{\!-\!1}$are strictly positive and even in $\omega$, and the FD ratio 
\beq
R_{q_x\omega}={N_1{\gamma} \over k_BT}
\left[1+\frac{\alpha\left(\alpha{N_2}/N_1+\beta\right)
{\left(\zeta/\!K\right)}{}^2q_x^2\Sigma(q_{\omega})}
{1-\alpha\beta{\left(\zeta/\!K\right)}{}^2q_x^2\Sigma(q_{\omega})}\right].
\label{NoneqFDT}
\eeq

At thermal equilibrium ($W\!=\!0$) 
$N_1=k_BT/\gamma$, $N_2=k_BT/\zeta$, $\alpha=A/\gamma$, and $\beta=-A/\zeta$, so that 
$\alpha{N_2}/N_1=-\beta$. Hence, the second term in the bracket in Eq.\ref{NoneqFDT} vanishes
and the fluctuation-dissipation ratio becomes unity as expected. With activity the
ratio becomes a strong function of frequency and wavenumber, through the quantity $\Sigma(q_{\omega})$. 
In the stiffening case  $\alpha\beta=(A/\gamma-Wc_0/\Gamma)(-A/\zeta)\sim{WA}\!>\!0$, if the strength 
$W$ of activity is large enough, as $\omega$ crosses a threshold which depends on $q_x$ and $W$, we see from Eqs.\ref{respimag},\ref{NoneqFDT}, $\chi{\prime\prime}$ can pass through zero, and hence $R_{q_x\omega}$ can diverge. Past this threshold, both turn negative. Thus one obtains a giant or even a negative FD ratio as a result of suppressed dissipation rather than enhanced noise.  The FD ratio Eq.\ref{NoneqFDT} is sometimes referred to \cite{hudspeth} as the ratio of an ``effective temperature'' $T_{eff}$ to the thermodynamic temperature $T$, and a negative $T_{eff}$ in this sense has been observed in experiments on hair cells \cite{hudspeth}. Our theory Eqs.\ref{ueqn}-\ref{veqn}, rooted in the active-hydrodynamic approach of \cite{RamaswamyActiveRheo1,curiegrp1,RamaswamyActiveRheo2,LM031,LM032,SriramMadanNJP07}, provides a fundamental basis for the model presented in \cite{hudspeth} and suggests that the phenomenon of negative dissipation should be widely observed in active systems. To apply our treatment directly to axons \cite{pramod2007} and hair cells \cite{hudspeth}, we consider a filament confined to a tube of radius $L$ in the $\perp$ directions and unbounded along $x$, and assume the correlation length $\xi \equiv \sqrt{K/a}$ of the active medium is of order $L$. The integrals over $\qp$ in Eq.\ref{QeqnFT} are replaced by sums dominated by a single mode with wavenumber of order $1/L$. This yields an effective equation of motion 
\beq
\label{ueqnsinglemode}
\left(-i\omega+{\sigma \over \gamma}q_x^2+{\kappa \over \gamma}q_x^4\right)
\up
= -q_x^2{\alpha \beta \zeta \over K}{1 \over 1 - {i \zeta L^2 \over K} \omega + q_x^2L^2} \up
\eeq
plus noise. For strong active stiffening, $\alpha \beta$ large and positive, this leads to oscillatory modes with $\omega \simeq \pm \sqrt{\alpha \beta}q_x/L$. These correspond to the hair cell oscillations of \cite{hudspeth}, generalized to allow a continuum of modes labelled by $q_x$. 

\section{Conclusion} 

We have presented a general theory for a single, long semiflexible filament interacting with an active medium. The medium consists of orientable elements, endowed with built-in uniaxial stresses whose axes are correlated with those of their neighbours. The model thus describes, in particular, a single microtubule interacting with an actively contractile actomyosin environment, but has broader applicability. Whereas the various components of the cytoskeleton are usually investigated separately, our paper lays the physical groundwork for a treatment of the interaction of these different elements, in particular the contractile actin, and microtubules as active transport highways. This is essential to an understanding of cell motility and active ribosomal transport. 

The key ingredients of the model are anchoring -- the preferred orientation imposed on the filaments of the medium when confronted with the surface of the long filament -- and the contractile or tensile activity of the medium. We find that the interplay between anchoring and activity radically affects the filament's dynamics, leading to a range of possible behaviours including active stiffening, negative dissipation, oscillations, and buckling. Our theory also applies to the dynamics of axons and auditory hair cells, and provides an important link between existing phenomenological models for these systems and the general framework of active matter. Our experiments on the shapes of microtubules in the growth cone of a neuron show both stiffening and buckling, in broad accord with our theory. We look forward to more quantitative tests, on filaments suspended in cell extracts as well as suspensions of swimming organisms, to see if the scenarii we propose are observed. 

\begin{acknowledgments}
NK was supported by the IISc Centenary Postdoctoral Fellowship.  SR
acknowledges support from the Department of Science and Technology, India
through the Centre for Condensed Matter Theory and DST Math-Bio Centre grant
SR/S4/MS:419/07. MR thanks the HFSP, and SR and MR thank CEFIPRA project
3504-2. We thank P. A. Pullarkat, J.-F. Joanny, J. Prost, N. Uchida, H. Ogawa
and T. Kawakatsu for stimulating discussions.  
\end{acknowledgments}
%%%%%%%%%%%%%%%%%%%%%%%%%%%%%%%%%%%%%%%%%%%%%%%%%%%%%%%%%%%

%%%%%%%%%%%%%%%%%%%%%%%%%%%%%%%%%%%%%%%%%%%%%%%%%%%%%%%%%%%%%%%

\end{document}